\documentclass{ws-procs10x7}
\usepackage{balance}

\usepackage{graphicx}

\newcommand{\epem}     {\ensuremath{\mathrm{e^+e^-}}}
\newcommand{\mpmm}     {\ensuremath{\mu^+\mu^-}}
\newcommand{\roots}    {\ensuremath{\sqrt{s}}}
\newcommand{\znull}    {\ensuremath{\mathrm{Z^0}}}
\newcommand{\mz}       {\ensuremath{m_{\znull}}}
\newcommand{\my}       {\ensuremath{m_{\mathrm{Y}}}}

\newcommand{\chisqd}   {\ensuremath{\chi^2/\mathrm{d.o.f.}}}

\newcommand{\as}     {\ensuremath{\alpha_{\mathrm{S}}}}

\newcommand{\asmz}   {\ensuremath{\as(\mz)}}

\newcommand{\ppbar}  {\ensuremath{\mathrm{p\bar{p}}}}
\newcommand{\bbbar}  {\ensuremath{\mathrm{b\bar{b}}}}
\newcommand{\qqbar}  {\ensuremath{\mathrm{q\bar{q}}}}

\newcommand{\sigl}   {\ensuremath{\sigma_{\mathrm{L}}}}

\newcommand{\ftwoep} {\ensuremath{F_2^{\mathrm{ep}}}}
\newcommand{\fthreenu} {\ensuremath{F_3^{\nu}}}
\newcommand{\ftwog}  {\ensuremath{F_2^{\gamma}}}
\newcommand{\gonen}  {\ensuremath{g_1^{\mathrm{N}}}}

\newcommand{\rhad}   {\ensuremath{R_{\mathrm{had}}}}
\newcommand{\rz}     {\ensuremath{R_{\znull}}}
\newcommand{\rtau}   {\ensuremath{R_{\tau}}}
\newcommand{\ryggg}  {\ensuremath{R_{\mathrm{Y\rightarrow\gamma gg}}}}

\newcolumntype{d}[1]{D{.}{.}{#1}}

\makeindex
\begin{document}

\title{REVIEW OF \as\ MEASUREMENTS}

\author{S. KLUTH}

\address{Max-Plank-Institut f\"ur Physik, D-80805 Munich, Germany\\
E-mail: skluth@mppmu.mpg.de}

\twocolumn[\maketitle\abstract{We present a review of measurements
of \as.  The individual measurements are discussed and intermediate
averages for classes of related measurements are found.  The final
average is built using the intermediate values.  Correlations are
treated consistently.  The ICHEP 2006 world average is
$\asmz=0.1175\pm0.0011$ dominated by the recent result from lattice
QCD.  }
]

\section{Introduction}
\label{sec_intro}

QCD is a highly predictive theory, because all processes involving
strong interactions must be described by QCD with a universal
parameter, the strong coupling constant \as.  Vice versa, all
determinations of \as\ using different processes must yield the same
results once different momentum scales or renormalisation schemes are
taken into account.  A systematic comparison of measurements
of \as\ thus is a strong test of the theory.

Averages of different measurements of \as\ are calculated with proper
treatment of correlations from common uncertainties~\cite{kluth06}.
Uncertainties are classified as statistical, experimental, soft QCD or
hard QCD.  Statistical and experimental uncertainties stem from
limited data samples and experimental systematics.  Soft QCD
uncertainties stem from hadronisation correction systematics, higher
twist effects, influence of parton density functions (pdfs) and other
non-perturbative effects.  Hard QCD uncertainties arise from unknown
or incomplete higher order corrections.

All results, intermediate averages and the final average are
shown in table~\ref{tab_as}.  The intermediate averages of \as\
from related analyses are found assuming uncorrelated statistical
errors, partially correlated experimental and soft QCD
errors and fully correlated hard QCD
errors~\cite{kluth06}.  Results with total error on \asmz\
significantly larger than 0.01 have not been considered.

\section{Lattice QCD}

New implementations of unquenched lattice QCD (LQCD) with dynamical
staggered light quarks (u, d and s) improve significantly the
description of some low energy observables~\cite{davies03} after
tuning the simulation with precisely known hadron masses and mass
differences.  Due to quark staggering quark vacuum polarisation loops
contribute 4-fold and the procedure has to be modified by hand to
compensate this effect.  The tuned LQCD is used to predict 28 selected
short distance observables which in turn are compared with NNLO QCD
calculations to extract \as~\cite{mason05}.  The uncertainties of the
result are dominated by limited simulation statistics and systematic
uncertainties of the analysis.  With this measurement a 1\% accuracy
for a determination of \asmz\ is reached.

\section{DIS Processes}

The analyses of scaling violation of structure functions (SFs) in deep
inelastic scattering (DIS) of leptons on nucleons result in precision
measurements of \as.  The scaling violation of the SF \ftwoep\ in e-p
DIS was studied in moment space in NNLO QCD but lacks a full analysis
of the theoretical error~\cite{santiago01}.  Following~\cite{bethke02}
the theory error is doubled.  The SF \fthreenu\ for neutrino-nucleon
DIS was analysed using Mellin moments and Jacobi polynomials in NNLO
QCD~\cite{kataev01}.  The SF \gonen\ for DIS with polarised
nucleons N was analysed in NLO QCD~\cite{bluemlein02}.

QCD predictions for some sum rules (SRs) in DIS are available in NNLO.
The study of the Bjorken SR for polarised DIS~\cite{ellis95} yields a
precise value of \as\ but the analysis of the experimental error has
been criticised~\cite{altarelli96,knauf02}; we double this error for
our averages.  The GLS~SR for $\nu$-N DIS was studied using CCFR
data~\cite{kim98}.

The more recent determinations of \as\ from jet production in e-p DIS
(J. Terron, these proceedings) are NLO QCD analyses covering a wide
range of $Q^2$ values.  The comprehensive combination of HERA
results for this process is used~\cite{glasman05}.

\section{Y Decays}

The Y resonances are \bbbar\ bound systems with mass dominated
by the large b quark mass.  Properties of these systems are
predicted with non-relativistic QCD (NRQCD) which takes the 
low velocities of the heavy quarks as an additional expansion
parameter.  Moments of $R_b(s=\my^2)=\sigma(\epem\rightarrow\bbbar)/
\sigma(\epem\rightarrow\mpmm)$~\cite{penin98} and the branching ratio
$\ryggg=\Gamma(\mathrm{Y}\rightarrow\gamma+\mathrm{hadrons})/
\Gamma(\mathrm{Y}\rightarrow\mpmm)$~\cite{pdg04} 
have been predicted in NRQCD and used to extract \as.

\section{\rtau}

For hadronic decays of $\tau$ leptons the invariant mass $s$ of the
hadronic final state sets the energy scale for QCD processes.  The
hadronic branching ratio
$\rtau(s)=\Gamma(\tau\rightarrow\nu_{\tau}\mathrm{hadrons})
/\Gamma(\tau\rightarrow\nu_{\tau}\nu_{\ell}\ell)$ is predicted in NNLO
QCD while non-perturbative effects are treated with the operator
product expansion (OPE).  Using $\tau$ decay data from~\cite{pdg04}
the determination of \as\ is updated~\cite{kluth06}.  A recent
analysis using the partially calculated NNNLO term is consistent but
claims smaller uncertainties~\cite{davier05}.

\section{Z Lineshape}

The precise data collected by the LEP experiments and SLD around the
\znull\ resonance yield an accurate determination of \asmz\ via QCD
corrections to electroweak processes.  The analysis~\cite{kluth06}
with data from~\cite{zedometry05} uses as observables the hadronic
width $\Gamma_h$ and the hadronic branching ratio
$\rz=\Gamma(\znull\rightarrow\mathrm{hadrons})
/\Gamma(\znull\rightarrow\ell\bar{\ell})$ of the \znull\ and the
on-peak hadronic and leptonic cross sections.  The result is
consistent with~\cite{zedometry05} and has a more complete error
analysis.

\section{\ftwog}

The scaling violations of the SF \ftwog\ for hadron production in
two-photon interactions at \epem\ colliders has been studied with NLO
QCD.  With recent data from LEP at high and low $Q^2$ a stable result
with small errors is obtained~\cite{albino02}.

\section{\rhad}

The analysis of hadron production in \epem\ annihilation at low
$\roots<2$~GeV uses the observable
$\rhad(s)=\sigma(\epem\rightarrow\mathrm{hadrons})/
\sigma(\epem\rightarrow\mpmm)$.  The NNLO analysis~\cite{menke01}
yields a fairly accurate result with errors dominated by experimental
uncertainties.

\section{\epem\ Jets and Event Shapes}

The determination of \as\ from jet rates and event shape distributions
is reviewed by J. Schieck in these proceedings.  The LEP experiments
have coordinated their final event shape analyses via a working group
(LEPQCDWG) yielding directly comparable consistent
results~\cite{kluth06}.  The re-analysis of JADE data uses the methods
developed at LEP and thus the results can also be compared
directly~\cite{kluth06}.  All analyses including the older TOPAZ
study~\cite{topaznlla} are based on NLO QCD calculations combined with
resummed NLLA calculations leading to more stable results.  The
availability of NLO QCD predictions combined with resummed NLLA
calculations for the 4-jet fraction with the Durham or Cambridge jet
algorithms lead to precise measurements of \as~\cite{jader4,kluth06}
from the LEP experiments and from JADE data.

\section{\epem\ Fragmentation}

In~\cite{kluth06} data from LEP experiments were used to update the
measurement of \as\ from the cross section \sigl\ for hadron
production via a longitudinally polarised virtual \znull\ or $\gamma$.
A combination of results for \as\ from analyses of the scaling
violation of charged hadron momentum spectra was done
in~\cite{kluth06}.  The data are from LEP and lower energy
experiments.  Both measurements are based on NLO QCD.

\section{pp/\ppbar\ Scattering Processes}

Both results stem from analysis of \ppbar\ collisions by the CERN
S\ppbar S collider experiments UA1 and UA6.  The cross section for
production of final states with b-jets is determined with a cut on the
angle between the two b-jets~\cite{ua1asbbbar95}.  Due to this cut the
cross section measurement becomes sensitive to \as; a NLO QCD
prediction is used to extract \as.

The cross section difference $\sigma(\ppbar\rightarrow\gamma\mathrm{X})-
\sigma(\mathrm{pp\rightarrow\gamma X})$ is sensitive to the parton
process $\qqbar\rightarrow\gamma\mathrm{g}$.  Together with DIS data
to constrain the valence quark pdfs \as\ can be
determined~\cite{werlen99}.

\section{ICHEP 2006 World Average}

\begin{figure}[htb!]
\centerline{
\includegraphics[width=\columnwidth]{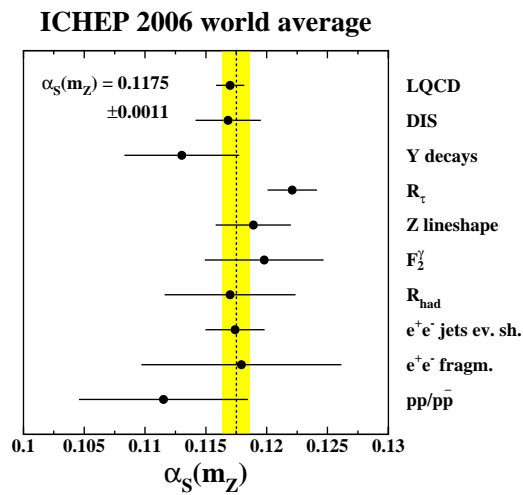}
}
\caption{ Intermediate averages of \asmz\ from each class of
analyses as shown in table~\ref{tab_as}.  The dashed vertical line and
grey band indicate the final average with total errors.  }
\label{fig_as}
\end{figure}

The final ICHEP 2006 world average is calculated from the values for
\asmz\ shown in table~\ref{tab_as} for each class of analyses.  In
case of several analyses in a class the intermediate average as shown
in table~\ref{tab_as} is used.  The statistical, experimental and soft
QCD errors are assumed to be uncorrelated and the hard QCD errors
are assumed to be partially correlated.  The intermediate and final
averages are shown in figure~\ref{fig_as}.  The final average is
dominated by the LQCD result with $\chisqd=17/9$ and $P(\chi^2)=0.05$.
Without the LQCD result the average is $\asmz=0.1200\pm0.0019$ with
$\chisqd=14/8$ and $P(\chi^2)=0.07$.  Our average is consistent with
other recent results~\cite{pdg04,bethke06}.  The small values for the
$\chi^2$ probabilities might indicate that the systematic errors in
some of the analyses are estimated aggressively.

\begin{table*}[htb!]
\tbl{ Results for \asmz\ from various analyses as indicated.  The values
of $Q$ refer to the energy scales were the measurements were performed.
\label{tab_as}}
{
\begin{tabular}{lcllllll}
\toprule
Process & $Q$ [GeV] & Theory & \asmz & $\pm$stat. & $\pm$exp. & $\pm$soft & $\pm$hard \\
\colrule
Lattice QCD~\cite{mason05} & 1.5-7.5 & NNLO & 0.1170 & 0.0007 & - & 0.0002 & 0.0009 \\
\colrule
\ftwoep~\cite{santiago01,bethke02} & 1.87-15.2 & NNLO & 0.1166 & -      & 0.0009 & 0.0006 & 0.0018 \\
pol. SF~\cite{bluemlein02} & 91.2 & NLO  & 0.113  & 0.004  & -      & 0.004  & 0.007 \\
\fthreenu~\cite{kataev01} & 2.2-11 & NNLO & 0.119  & 0.002  & 0.005  & 0.001  & 0.004 \\
Bjorken SR~\cite{ellis95} & 1.58 & NNLO & 0.1217 & 0.0026 & 0.0092 & 0.0029 & 0.0006 \\
GLS SR~\cite{kim98} & 1.73 & NNLO & 0.1123 & 0.0048 & 0.0066 & 0.0045 & 0.0006 \\
$ep\rightarrow$jets~\cite{glasman05} & 8.5-60 & NLO & 0.1186 & - & 0.0011 & - & 0.0050 \\
DIS av. & - & -    & 0.1168 & 0.0002 & 0.0009 & 0.0006 & 0.0025 \\
\colrule
$R_b(s)$~\cite{penin98} & 4.75 & NNLO & 0.1191 & -      & 0.0038 & -      & 0.0038 \\
\ryggg~\cite{pdg04} & 4.75 & NNLO & 0.1097 & - & 0.0032 & - & 0.0032 \\
Y decays av. & - & - & 0.1130 & - & 0.0033 & - & 0.0034 \\
\colrule
\rtau~\cite{kluth06} & 1.777 & NNLO & 0.1221 & - & 0.0006 & 0.0004 & 0.0019 \\
\colrule
\znull\ Lineshape~\cite{kluth06,zedometry05} & 91.2 & NNLO & 0.1189 & - & 0.0027 & - & 0.0015 \\
\colrule
\ftwog~\cite{albino02} & 1.38-27.9 & NLO & 0.1198 & - & 0.0028 & - & 0.0040 \\
\colrule
\rhad~\cite{menke01} & 2 & NNLO & 0.117 & - & 0.005 & - & 0.002 \\
\colrule
JADE ev. sh.~\cite{kluth06} & 14-44 & NLO+NLLA & 0.1203 & 0.0007 & 0.0017 & 0.0053 & 0.0050 \\
TOPAZ ev. sh~\cite{topaznlla} & 58 & NLO+NLLA & 0.1219 & 0.0036 & 0.0008 & 0.0015 & 0.0047 \\
LEP ev. sh.~\cite{kluth06} & 91.2-206 & NLO+NLLA & 0.1202 & 0.0005 & 0.0008 & 0.0019 & 0.0049 \\
JADE $R_4$~\cite{jader4} & 14-44 & NLO+NLLA & 0.1159 & 0.0004 & 0.0012 & 0.0024 & 0.0007 \\
LEP $R_4$~\cite{kluth06} & 91.2-206 & NLO(+NLLA) & 0.1175 & 0.0002 & 0.0010 & 0.0014 & 0.0015 \\
\epem\ jets ev. sh. av. & - & - & 0.1174 & 0.0002 & 0.0010 & 0.0015 & 0.0016 \\
\colrule
\epem\ \sigl~\cite{kluth06} & 91.2 & NLO & 0.1169 & - & 0.0035 & 0.0018 & 0.0072 \\
\epem\ sc. viol.~\cite{kluth06} & 91.2-206 & NLO & 0.1192 & - & 0.0056 & - & 0.0070 \\
\epem\ fragm. av. & - & - & 0.1179 & - & 0.0040 & 0.0010 & 0.0071 \\
\colrule
$\ppbar\rightarrow\bbbar X$~\cite{ua1asbbbar95} & 20 & NLO & 0.1130 & - & 0.0065 & 0.0050 & 0.0067 \\
$pp/\ppbar\rightarrow\gamma X$~\cite{werlen99} & 24.3 & NLO & 0.1112 & 0.0016 & 0.0033 & 0.0039 & 0.0039 \\
pp/\ppbar\ av. & - & - & 0.1115 & 0.0013 & 0.0034 & 0.0039 & 0.0044 \\
\colrule
ICHEP 2006 av. & - & - & 0.1175 & 0.0006 & 0.0001 & 0.0002 & 0.0009 \\
\botrule
\end{tabular}
}
\end{table*}

\end{document}